\documentclass{PoS}

\newcommand{\be}{\begin{equation}}
\newcommand{\ee}{\end{equation}}
\newcommand{\bea}{\begin{eqnarray}}
\newcommand{\eea}{\end{eqnarray}}
\newcommand{\mbb}{\mathbb}
\newcommand{\mc}{\mathcal}
\newcommand{\vo}{\mathcal{V}}
\newcommand{\nn}{\nonumber}

\title{Large extra dimensions and light hidden photons from anisotropic string vacua}

\ShortTitle{Large extra dimensions and light hidden photons from string vacua}

\author{\speaker{Michele Cicoli}\thanks{The results described here have been found
in collaboration with C. Burgess, M. Goodsell, J. Jaeckel, F. Quevedo and A. Ringwald.}\\
        Abdus Salam ICTP, Strada Costiera 11, Trieste 34014, Italy\\
        E-mail: \email{mcicoli@ictp.it}}

\abstract{We derive type IIB vacua which are very promising to put string theory
to experimental test. These are Calabi-Yau compactifications
with a 4D fibration over a 2D base. The moduli are fixed
in such a way to obtain a very anisotropic configuration
where the size of the 2D base
is exponentially larger than the size of the 4D fibre.
These provide stringy realisations
of the supersymmetric large extra dimensions scenario
and extensions of the ADD scenario which are characterised
by TeV-scale strings and two micron-sized extra dimensions.
We also study the phenomenological properties of hidden Abelian gauge bosons
which mix kinetically with the ordinary photon and get a mass via the Green-Schwarz mechanism.
We show that anisotropic compactifications lead naturally to
dark forces for an intermediate string scale or even to a hidden CMB
for the extreme case of TeV-scale strings.}

\FullConference{The 2011 Europhysics Conference on High Energy Physics-HEP 2011,\\
		July 21-27, 2011\\
		Grenoble, Rhones Alpes France}

\begin{document}

\section{Micron-sized extra dimensions from anisotropic compactifications}

Type IIB flux compactifications are very promising to test string vacua.
In particular, LARGE Volume Scenarios (LVS) can naturally give rise to strings at LHC scales
and two micron-sized extra dimensions (EDs) for compactifications on fibred Calabi-Yau (CY) three-folds \cite{ADDstrings}.
The internal volume $\vo$ is fixed exponentially large in string units:
$\vo \simeq e^{c/g_s}\gg 1$, where $c$ is an $\mc{O}(1)$ parameter and the string coupling
$g_s\ll 1$ is in the perturbative regime. $M_s\simeq 1$ TeV can be obtained for $\vo\simeq 10^{30}$
since dimensional reduction gives $\vo \simeq M_p^2/M_s^2$,
providing a dynamical solution of the hierarchy problem based on moduli fixing.
If all the EDs are of the
same size $L\simeq \vo^{1/6} M_s^{-1} \simeq \left(10\, {\rm MeV}\right)^{-1}\simeq 10 \,{\rm fm}$,
$L$ is much smaller than the current bounds from tests of Newton's law.
Thus if we want to obtain a much larger $L$, we need to
find anisotropic solutions with $\vo \simeq L^2 l^4 M_s^6$ where $L\simeq 10 \,\mu {\rm m}\simeq\left(1\,{\rm meV}\right)^{-1}$
and $l\simeq 10^{-4} \,{\rm fm} \simeq \left(1 \,{\rm TeV}\right)^{-1}\ll L$,
that would lead to a stringy derivation of ADD scenarios with two micron-sized EDs \cite{ADD}.
This is why we focus on CY three-folds with a 2D $\mbb{P}^1$ base with volume $t_1 := \left(L M_s\right)^2$,
a 4D K3 or $T^4$ fibre with size $\tau_1 := \left(l M_s\right)^4$
and a 4D del Pezzo with volume $\tau_3 := \left(d M_s\right)^4$. In fact,
the CY volume $\vo\simeq t_1\tau_1- \tau_3^{3/2}$ \cite{ToricCY},
in the limit $ t_1\tau_1\gg \tau_3^{3/2}$, simplifies to
the desired form $\vo\simeq t_1\tau_1 \simeq L^2 l^4 M_s^6$.
Given that in LVS the leading dynamics fixes only
the combination of $t_1$ and $\tau_1$ corresponding to $\vo\simeq 10^{30}$,
one flat direction is left over. Hence, if we parameterise it by $\tau_1$,
we need to fix $\tau_1\simeq \mc{O}(10)$ so that
$\langle t_1\rangle \gg \sqrt{\langle\tau_1\rangle}\simeq\sqrt{\langle\tau_3\rangle}$
$\Rightarrow$ $L\gg l\simeq d$.

\medskip{\bf Moduli stabilisation}:
Due to the no-scale structure, the K\"ahler moduli are fixed beyond leading order in $\alpha'$ and $g_s$.
The $\alpha'$ corrections to the K\"ahler potential $K$
do not develop any potential for $\tau_1$ since they depend only on $\vo$.
Open string loop corrections to $K$ depend on $\tau_1$
if there is a D7-stack wrapping the fibre. In this case,
they would fix $\tau_1$ too large at $\langle\tau_1\rangle \simeq g_s^{4/3}\,\langle\vo\rangle^{2/3}\gg 1$.
If there are instead no D7-branes wrapping $\tau_1$,
we expect no $\tau_1$-dependent $g_s$ correction to $K$ since open strings are
localised far away from the fibre. In this case, $\tau_1$ can be fixed only
by non-perturbative corrections to the superpotential $W$.
In the presence of a racetrack superpotential on $\tau_3$, $W=W_0+A\, e^{-a_3 T_3}- B\, e^{-b_3 T_3}$,
the del Pezzo divisor can be fixed at $\langle\tau_3\rangle\simeq 1/g_s\simeq \mc{O}(10)$ and
the volume at $\langle\vo\rangle\simeq e^{c/g_s}\simeq 10^{30}$. The fibre is then fixed
at $\langle\tau_1\rangle \simeq \langle\tau_3\rangle\simeq \mc{O}(10)$
via poly-instanton corrections to $W$ generated by a string instanton on $\tau_1$:
\be
    W = W_0 + A \,e^{-a_3\left(T_3+C_1 e^{-2\pi T_1}\right)}-B \,e^{-b_3\left(T_3+C_2 e^{-2\pi T_1}\right)}\,. \nn
\ee
The poly-instanton potential for $\tau_1$,
$V_{\rm poly} \simeq \mc{O}(\vo^{-4})$, is suppressed since the
leading potential scales as $V_{\rm lead} \simeq \mc{O}(\vo^{-3})$. Closed string
loops are $\tau_1$-dependent but they can be shown to be negligible.

\medskip{\bf Mass scales}: The higher dimensional Planck scales are $M_{10D} \simeq M_s$
and $M_{6D} = \left(4 \pi\tau_1\right)^{1/4} M_s$, while the 4D Planck scale is $M_p = \sqrt{4 \pi \vo} \, M_s$.
There are also different Kaluza-Klein (KK) scales: the theory becomes 6D above
$M_{\rm KK}^{6D}\simeq M_s/t_1^{1/2}\simeq 1/L$ and 10D above $M_{\rm KK}^{10D} \simeq M_s/\tau_1^{1/4} \simeq 1/l$ while
KK replicas of Standard Model (SM) particles show up at $M_{\rm KK}^{\rm SM} \simeq M_s/\tau_{\rm SM}^{1/4}\simeq 1/d$,
where $\tau_{\rm SM}$ is an additional del Pezzo supporting the SM brane stack.
For $M_s\simeq 3$ TeV, we would have $M_{6D}\simeq 10$ TeV, $M_{10D}\simeq 4$ TeV,
$M_{\rm KK}^{\rm SM}\simeq M_{\rm KK}^{10D}\simeq 1$ TeV and $M_{\rm KK}^{6D}\simeq 1$ meV.
The dilaton and the complex structure moduli are almost degenerate with the gravitino which has
a mass $m_{3/2}\simeq M_s^2/M_p \simeq 1$ meV.
The K\"ahler moduli can be even lighter due to no-scale structure.
In fact, $\tau_3$ has a mass of order $m_{3/2}$ but $\tau_1$ and $\vo$ are much lighter
since $m_1 \simeq M_p/\vo^2\simeq 10^{-32}$ eV and $m_{\vo} \simeq M_p/\vo^{3/2}\simeq 10^{-18}$ eV.
These moduli would mediate unobserved fifth forces but we still need to study radiative corrections.

\medskip{\bf Supersymmetry breaking}: Compactifications with TeV-scale strings need large supersymmetry (SUSY) breaking
effects since gravity mediation would yield soft terms of order $M_{\rm soft}\simeq m_{3/2}\simeq 1$ meV.
Hence we consider a brane set-up that breaks SUSY explicitly,
so that the low-energy theory contains just the SM.
The bulk is instead approximately supersymmetric since $m_{3/2}\simeq 1$ meV,
providing a stringy realisation of supersymmetric large EDs scenarios
which are promising for explaining dark energy \cite{SLED}.
Given that SUSY is badly broken on the SM brane, we expect
large radiative corrections to moduli masses from loops of massive open strings:
\be
\delta m \simeq \zeta \, M_s^2/M_p \simeq \zeta \, M_p/\vo \quad {\rm where}\quad
\mc{L}_{\rm int} = \left(\zeta/M_p\right) \,\phi \, F_{\mu\nu} F^{\mu\nu}\,, \nn
\ee
for a canonically normalised modulus $\phi$. In the case of $\tau_1$, $\zeta\simeq 1/\vo$,
as this is a flat direction at leading order. Hence
the mass of the fibre is unchanged. However, since this modulus is very weakly coupled,
$g \simeq 1/(M_p \vo)$, there are no bounds from fifth forces. For the volume mode,
$\zeta \simeq 1$, and so its mass gets lifted to $m_{\vo} \simeq 1$ meV,
just at the edge of detectability in fifth force experiments.

\medskip{\bf Phenomenology}: The phenomenology of these models differs from ADD scenarios due to the
presence of new states and low-energy bulk SUSY. There are many
implications for colliders, astrophysics and cosmology:
$(i)$ deviations from Newton's law;
$(ii)$ energy loss into the EDs due to radiation of KK modes;
$(iii)$ absence of superpartners of SM particles
since SUSY is non-linearly realised;
$(iv)$ strong bounds from neutron-star cooling ($M_{6D} > 700$ TeV)
and distortions of the MeV diffuse $\gamma$-ray background
can be evaded since KK modes decay into invisible degrees of freedom (\emph{dof});
$(v)$ problems with BBN, if the SM brane cools too quickly through evaporation into the bulk,
require a very low reheat temperature $T_{\rm RH}\lesssim 100$ MeV;
$(vi)$ no over-closure of the Universe since the lightest KK mode cannot be stable
due to the absence of continuous isometries in a compact CY;
$(vii)$ the lightest moduli and KK modes are good dark matter candidates.

\section{Light hidden photons from anisotropic compactifications}

Hidden photons $\gamma'$ interact
with the visible sector \emph{only} via kinetic mixing with the hypercharge:
\be
\mc{L} \supset  -\frac{1}{4} F^{({\rm vis})}_{\mu \nu} F^{\mu \nu}_{({\rm vis})}
- \frac{1}{4} F^{({\rm hid})}_{\mu \nu} F^{\mu \nu}_{({\rm hid})}+  \frac{\chi}{2} F_{\mu \nu}^{({\rm vis})} F^{({\rm hid}) \mu \nu}
+ m_{\gamma'}^2 A^{({\rm hid})}_{\mu} A^{({\rm hid})\mu}
+A^{({\rm vis})}_{\mu}j^{\mu}\,. \nn
\ee
The hidden photon can get a mass $m_{\gamma'}$ either via a model-dependent Higgs mechanism in the hidden sector
or via a St\"uckelberg mechanism which is typically stringy since it corresponds to the
Green-Schwarz cancellation of the anomalies. Here we shall focus on the St\"uckelberg mechanism.
The kinetic mixing takes place at one-loop
and the corresponding parameter scales as $\chi \simeq g_Y g_{\rm hid}/\left(16\pi^2\right)$.

Similar to neutrinos, the kinetic mixing induces photon $\leftrightarrow$ hidden photon oscillations.
Given that thermal photons get a plasma mass $\omega_P \simeq 1$ meV,
a resonant conversion of photons into $\gamma'$ with $m_{\gamma'}\simeq 1$ meV can occur after BBN but before CMB
decoupling. This results in a \emph{hidden CMB} which increases the effective number of relativistic \emph{dof}.
The current observational value $\Delta N_\nu^\mathrm{eff}=1.3\pm 0.9$
can be obtained for $\chi\simeq 10^{-6}$ in a region that is tested experimentally.
Moreover, due to the kinetic mixing, SM particles get a small charge under the hidden $U(1)$ leading to \emph{dark forces}.
A mass of the order $m_{\gamma'}\simeq 1$ GeV can explain either
the deviation of $(g-2)_{\mu}$ from the SM prediction if $\chi\simeq 10^{-3}\div 10^{-2}$,
or several puzzling observations connected to dark matter and astrophysics if $\chi\gtrsim 10^{-6}$,
again in a region tested by experiments. For a review on these issues see \cite{HiddenReview}.

\medskip{\bf Hidden photons as open strings}: Hidden photons arise naturally
in string compactifications. In type IIB vacua, the $\gamma'$ is an excitation of a D7-brane
wrapping a four-cycle $\tau_{\rm hid}$ far from the SM. For large $\tau_{\rm hid}$,
the hidden gauge coupling becomes very small since $g^{-2}_{\rm hid}= \tau_{\rm hid}/(4\pi)\ll 1$
resulting in a significantly suppressed $\chi$ which can naturally be of order $10^{-6}$.
By turning on a world-volume flux on the hidden D7-stack,
a K\"ahler modulus $\tau_j$ gets a charge $q\neq 0$ under $U(1)_{\rm hid}$ and the corresponding
axion gets eaten up by the $\gamma'$ which acquires a St\"uckelberg mass
$m_{\gamma'}^2\simeq q^2 g_{\rm hid}^2 (K_0)_{jj} M_p^2$,
where $(K_0)_{jj}$ is the $jj\,$ element of the tree-level K\"ahler metric.
Moreover, a moduli-dependent Fayet-Iliopoulos term gets generated.
As can be seen from the expression for $m_{\gamma'}$, if the EDs are exponentially large,
hidden photons can naturally be very light \cite{GJRR}.
In the case of isotropic compactifications, no prediction
can be obtained in the interesting regions of the $(\chi, m_{\gamma'})$ parameter space \cite{GJRR}.
However anisotropic compactifications with hidden photons
living on D7-branes wrapping the fibre divisor, predict a \emph{hidden CMB} and \emph{dark forces}
which can be detectable in the lab \cite{HiddenPhotons}.
Furthermore, all the moduli can be successfully fixed with the
D-terms under control \cite{HiddenPhotons}.

\medskip{\bf Phenomenological implications}:
Let us focus on LVS anisotropic compactifications with $\gamma'$ living on the fibre divisor.
$\tau_1$ is fixed via $g_s$ corrections to $K$ at $\langle \tau_1 \rangle =  \kappa \,\langle\tau_2\rangle$ where
$\kappa\simeq\left(g_s c_1\right)^2/c_2$ with $c_1$ and $c_2$ unknown coefficients of the loop corrections.
The relation between $m_{\gamma'}$ and $\chi$ can be written as $m_{\gamma'}\simeq \kappa \,10^{24} \chi^3$ GeV,
leading to two interesting scenarios:
\begin{enumerate}
\item \textbf{Natural dark forces for intermediate scale strings}:
$m_{\gamma'}\simeq 1$ GeV and $\chi\simeq 10^{-6}$ for $\kappa \simeq 10^{-6}$
which can be achieved without fine-tuning the underlying parameters.
The string scale is $M_s\simeq 10^{11}$ GeV and the CY is slightly anisotropic:
$L\simeq 10^4 \,M_s^{-1} >l\simeq 10^2\,M_s^{-1}$.

\item \textbf{Hidden CMB with KK dark forces and strings at the LHC}:
$m_{\gamma'}\simeq 1$ meV and $\chi\simeq 10^{-6}$ for $\kappa \simeq 10^{-18}$
which requires a severe fine-tuning and gives TeV-scale strings.
KK hidden photons with $M^{\gamma'}_{\rm KK}\simeq M_s \tau_1^{-1/4}\simeq 1$ GeV
might behave as dark forces and the CY is very anisotropic: $L\simeq 10^{11}\,M_s^{-1}\gg l\simeq 10^2\,M_s^{-1}$.
\end{enumerate}


\begin{thebibliography}{99}

\bibitem{ADDstrings}
M.~Cicoli, C.~P.~Burgess and F.~Quevedo,
\emph{Anisotropic Modulus Stabilisation: Strings at LHC Scales with Micron-sized Extra Dimensions},
[{\tt arXiv:1105.2107 [hep-th]}].

\bibitem{ADD}
N.~Arkani-Hamed, S.~Dimopoulos and G.~R.~Dvali,
\emph{The hierarchy problem and new dimensions at a millimeter},
\emph{Phys.\ Lett.\ B} {\bf 429} (1998) 263 [{\tt arXiv:hep-ph/9803315}].

\bibitem{ToricCY}
M.~Cicoli, M.~Kreuzer, C.~Mayrhofer,
\emph{Toric K3-Fibred Calabi-Yau Manifolds with del Pezzo Divisors for String Compactifications},
[{\tt arXiv:1107.0383 [hep-th]}].

\bibitem{SLED}
Y.~Aghababaie, C.~P.~Burgess, S.~L.~Parameswaran and F.~Quevedo,
\emph{Towards a naturally small cosmological constant from branes in 6D supergravity},
\emph{Nucl.\ Phys.\  B} {\bf 680} (2004) 389.

\bibitem{HiddenReview}
  J.~Jaeckel and A.~Ringwald,
\emph{The Low-Energy Frontier of Particle Physics},
\emph{Ann.\ Rev.\ Nucl.\ Part.\ Sci.}  {\bf 60} (2010) 405
  [{\tt arXiv:1002.0329 [hep-ph]}].

\bibitem{GJRR}
M.~Goodsell, J.~Jaeckel, J.~Redondo and A.~Ringwald,
\emph{Naturally Light Hidden Photons in LARGE Volume String Compactifications},
\emph{JHEP} {\bf 0911} (2009) 027 [{\tt arXiv:0909.0515 [hep-ph]}].

\bibitem{HiddenPhotons}
M.~Cicoli, M.~Goodsell, J.~Jaeckel, A.~Ringwald,
\emph{Testing String Vacua in the Lab: From a Hidden CMB to Dark Forces in Flux Compactifications},
\emph{JHEP} {\bf 1107} (2011)  114 [{\tt arXiv:1103.3705 [hep-th]}].

\end{thebibliography}
\end{document}